\documentclass[12pt]{article}
\usepackage{graphicx}
\usepackage{amsmath}
\usepackage{amssymb}
\usepackage{amsfonts}
\usepackage{caption}
\usepackage{mathtools}
\usepackage{color}
\usepackage[utf8]{inputenc}
\usepackage[english]{babel}
\usepackage[table]{xcolor}
\usepackage{setspace}
\usepackage{pdfpages}
\allowdisplaybreaks

\usepackage{array}
\usepackage{cellspace}
\usepackage[left=2.5cm,right=2.5cm, top=3cm, bottom=3cm]{geometry}
\newcolumntype{L}[1]{>{\raggedright\arraybackslash} m{#1} }
\usepackage[breaklinks=true,colorlinks=true,linkcolor=blue,urlcolor=blue,citecolor=blue]{hyperref}

\begin{document}
\title{\Huge Transformation equation for  frames undergoing non-uniform acceleration such as SHM  and  rotational motion}

  \author{\text{Ranchhaigiri Brahma$^{1\ast}$ } and \text{A.K. Sen$^{1\dagger}$  }\\
  \\ \textsl{$^1$Department of Physics, Assam University, Silchar-788011, India}\\ $^{\ast}$ Email: \href{mailto: rbrahma084@gmail.com}{\textit{ rbrahma084@gmail.com}}\\
  $^{\dagger}$ Email: \href{mailto: asokesen@yahoo.com}{\textit{ asokesen@yahoo.com}}}
\date{  }
\maketitle
%\doublespacing
\onehalfspacing
\abstract{Lorentz  transformation equations  provide us  a set of  relations between the spacetime  coordinates  as  observed  from two  different inertial  frames. In case, one of the frames is moving with a uniform rectilinear acceleration we have  Rindler's  transformation equations under such a situation. In the present work,  we extend the Rindler's  equations to a situation where  we have in general  non-uniform acceleration.   After that  we consider the  non-inertial  frame  to undergo simple  harmonic  motion (SHM)  and  as  a second case  we  consider  the  non-inertial frame to move uniformly along  a circle. This  set  of  transformation equations  will have  applications in various branches  of  Physics  and  in general in Astrophysics.}\\\\
\textbf{Keywords:} \textit{Relativity; Non-inertial frames; Simple harmonic motion (SHM); Uniform circular motion (UCM).}

\section{Introduction}
In special theory of relativity, the coordinate transformation relations for a frame of reference moving with uniform velocity with respect to a rest frame are known as the Lorentz transformation \cite{padmanbhan}. These relations play a vital role during the investigation of a test particle in a frame moving with a uniform velocity. 
In the case of an accelerated frame, the general transformation relations are presented by Nelson \cite{nelson1987generalized,1994JMP....35.6224N}. The coordinate transformations for uniformly accelerated frame with constant acceleration is known as Rindler transformation and the consequences of such coordinate frame have been studied by different researchers \cite{rindler1966kruskal, David1997, Yi, 2015GReGr..47...33D, Sanchez, boumali2021thermal}. In Rindler transformation the initial velocity is considered as zero. However the coordinate transformation for accelerated frame with non-zero initial velocity has been discussed by Yi \cite{Yi}. Additionally,  Brahma and Sen \cite{brahma2024gravitationalfieldmovingschwarzschild} analyzed the Rindler transformation having acceleration along an arbitrary direction on the XY plane. The  physics  with the accelerated frames  can  be also  very efficiently  handled  with the help of  Fermi coordinates  which  are local coordinates  adapted to any world line ( even if  it  is not geodesical) 
 \cite{10.1063/1.1724316}. Using  Fermi coordinates  which are basically  generalization of an inertial Cartesian coordinates to an accelerated system, Nesterov \cite{Nesterov_1999} analyzed the plane-wave metric for a geodesic observer in a weak gravitational-wave field. In another work, Chicone and Mashhoon \cite{Chicone_2004} also investigated the motion of ultra-relativistic particles and light rays using the Fermi coordinates. \\\\
 Scientists   also  use {\it proper reference frame} in the theory of relativity, which is a particular form of accelerated reference frame where an accelerated observer can be considered as being at rest.  Kajari et al. \cite{kajari_2009} discussed the propagation of light using {\it proper reference frame}.
Using the hypothesis of locality, Mashhoon and Muench \cite{Mashhoon_2002}  investigated the limitations of length measurements by accelerated observers in Minkowski
spacetime brought about via the hypothesis of locality, namely, the assumption that an accelerated observer
at each instant is equivalent to an otherwise identical momentarily comoving inertial observer. They found 
that the consistency can be achieved only in a rather limited neighborhood around the observer with linear dimensions that are negligibly small compared to the characteristic acceleration length of the observer. Furthermore, Nikoli\'c \cite{Nikoli__2000} analyzed the rotational motion in special relativity by calculating transformation relations from Nelson's general transformation relations \cite{nelson1987generalized, 1994JMP....35.6224N}.\\\\ 
In the present work, we investigate the coordinate transformation relations for a non-uniform acceleration such as a frame executing simple harmonic motion (SHM) and subsequently the transformation relations for a frame having uniform circular motion (UCM). 
However, in special relativity, when the motion of a test particle is subject to a restoring force along the $X-$direction, the mass of the particle can not be considered as constant. But, it changes according to the law of special theory of relativity to which MacColl \cite{1957AmJPh..25..535M} calculated a general expression of the time period of the relativistic oscillation, which is a function of the total energy of the system. Such relativistic harmonic oscillation can be used to admit a consistent procedure for the quantization of the system \cite{PhysRevD.6.1474}. However, the formal investigation on the harmonic oscillator in relativity using the Lagrange formalism \cite{10.1119/1.17513} and the Hamiltonian formalism \cite{PhysRevE.87.033202} indicates that even though the term ``harmonic oscillator" has been generally used, but the results obtained during the calculations are no longer harmonic in nature. Instead, it is ``anharmonic" in nature. It indicates that if the maximum velocity of the SHM is comparable to the speed of light, then the harmonic oscillation becomes an anharmonic in nature. The nature of the relativistic harmonic oscillation and the associated anharmonicity due to the relativistic effect have been experimentally studied using the ultracold lithium atoms in the third band of an optical lattice and reported by Fujiwara et al. \cite{Fujiwara_2018}. Here, it has been stated that the measured worldline shapes, relativistic anharmonicity, monopole oscillations, and relativistic dephasing of oscillator ensembles are in good agreement with the theoretical predictions of the relativistic harmonic oscillation. It is to be mentioned that, although the simple harmonic oscillation has been studied extensively in the regime of relativity, but only a few works on relativistic coordinate transformation between an inertial frame and a frame executing SHM have been reported in literature to our knowledge. For example, Perepelkin et al. \cite{Perepelkin_2016} discussed the transformation relation for a frame having velocity which is a sinusoidal function of time by using generalized Galilean transformation. Therefore, it would  be  very  much  important  and  appropriate, to study the transformation relations for the frame undergoing SHM through a more  general  procedure. This motivated us to calculate such coordinate transformation relations in the present work. \\\\
So in the present context, in order to carry out the calculations to obtain the coordinate transformations between an inertial frame and a frame executing  SHM, we have to first derive Rindler like transformation equations for coordinates frames having non-uniform acceleration. Since the circular motion of an object can be described by two SHMs with same amplitudes in perpendicular directions having phase difference of $\pi/2$ radians, so the transformation relations for the frame executing SHM can be used to understand the dynamics of a moving frame in a circular path. As a result, we obtain the coordinate transformation relations for a frame of reference moving in a circular path. The results in the present work will have immense applications in the study of the test particles in the non-uniformly accelerated frames such as frames executing SHM and the frames moving in a circular path.\\\\
We organized our paper as: In {\bf Section 2} we obtain the transformation relations for a frame of reference executing simple harmonic motion (SHM) along $X-$direction. Additionally, we also calculated the similar transformation relations for the frame of reference executing SHM along the $Y-$direction with initial phase of $\pi/2$ radians. And then in {\bf Section 3}, as mentioned above since the circular motion can be described by using the two SHMs in perpendicular direction and phase difference of $\pi/2$ radians, towards the end we obtained the transformation relations for a frame that is moving in a circular path.  In {\bf Section 4} we compared our results with the existing coordinate transformation relations for the frames executing SHM and UCM available in literature.  At last in {\bf Section 5} we discussed the results obtained in the above calculations and concluded our findings.

\section{Transformation relations for a frame of reference executing SHM}
The original Rindler transformation equation is meant for constant acceleration. However, in the present work, we shall derive the transformation relations in which proper acceleration or coordinate acceleration is a function of time. Here, we consider two inertial frames $S(ct, x,y,z)$ and $S'(ct', x', y',z')$, where the $S'$ frame is moving with uniform velocity $v$ with respect to $S$ frame along the $X-$direction. Then the coordinate velocity and acceleration of a particle along the $X-$direction in the $S$ frame can be expressed as $u_x=\frac{dx}{dt}$ and $a_x=\frac{du_x}{dt}$, respectively. Similarly, the coordinate velocity and acceleration of a particle along the $X$-direction in the $S'$ frame can be expressed as $u'_x=\frac{dx'}{dt'}$ and $a'_x=\frac{du'_x}{dt'}$, respectively. As available in the literature \cite{wRindler, misner, wRindler3},  the proper time is the  time   measured  by a clock tied to  the frame $S'$ given by $\tau= \int \sqrt {1-v^2/c^2}dt$.  Furthermore, the proper velocity and  proper acceleration are  defined as $dx/d\tau$ and  $d(dx/d\tau)/dt$ respectively.  We denote the  proper acceleration by $\alpha$ in our work.  Physically, proper  acceleration is that which is measured by an accelerometer carried along with the accelerated object's frame. Most specifically for a frame executing simple harmonic motion (SHM) the coordinate acceleration is a sinusoidal function of coordinate time in that frame. Below, we derive the transformation relations for the case when the frame is oscillating along the $X-$direction having displacement proportional to some sinusoidal function of angular frequency $\omega$ as  defined in the moving frame $S'$.\\\\
The two inertial frames $S$ and $S'$ will have the transformation law for acceleration as \cite{Sanchez}:
\begin{align}
    a'_x=\frac{\left(1-\frac{v^2}{c^2}\right)^{3/2}}{\left(1-\frac{u_x\,v}{c^2}\right)^{3}}a_x
\end{align}
where $u_x=\frac{dx}{dt}$ and $a_x=\frac{du_x}{dt}$ respectively are the coordinate velocity and the coordinate acceleration in the $S$ frame; $a'_x=\frac{du'_x}{dt'}=\frac{d^2x'}{dt'^2}$ represents the coordinate acceleration of a particle in the $S'$ frame. Therefore, for a particle that has acceleration $a'_x$ with respect to an inertial frame $S'$ which instantaneously accompanies the particle $(u_x=v)$, we can write:
\begin{align}
    &\frac{a_x}{\left(1-\frac{u_x^2}{c^2}\right)^{3/2}}=a^{\prime} _x \label{A5}
\end{align}
In case of  Castillo \& Sanchez \cite{Sanchez} they assumed that the acceleration $a^{\prime} _x $ is constant  and  obtained the  Rindler's transformation equation \cite{PhysRev.119.2082}. However, in our present case  we  shall consider $a^{\prime} _x $  to be  a function of  $(x^\prime, t^\prime)$  and  proceed  as follows.\\\\
From equation \eqref{A5} we can write,
\begin{align}
    &\frac{du_x/dt}{\left(1-\frac{u_x^2}{c^2}\right)^{3/2}}=a'_x\nonumber\\
    \text{or\,\,\,\,\,} &\frac{du_x}{\left(1-\frac{u_x^2}{c^2}\right)}=a'_x\,\left(1-\frac{u_x^2}{c^2}\right)^{1/2} dt\label{q3}
\end{align}
The quantity $u_x$ in the right hand side is actually representing $v$. So we can identify $\left(1-\frac{u_x^2}{c^2}\right)^{1/2} dt=d\tau$, where $\tau$ is the proper time. Additionally from equation \eqref{A5} we can also find, the acceleration $a_x'$ becomes  equal  to  the proper acceleration ($\alpha$),  because  we have taken $u_x=v$.  The details are  available in  \cite{wRindler3}. This is  because  $a_x'$ is  actually equivalent to the  differential  co-efficient of  proper  velocity  with respect  to  co-ordinate  time.  Also, since the proper time $\tau$ is inherently related to the coordinate time $t'$ in the $S'$ frame, so in the present context we can easily express the acceleration $a'_x$ as a function of ($\tau, x'$). Thus, we can finally write:
\begin{align}
    \frac{du_x(t, x)}{\left(1-\frac{u_x^2}{c^2}\right)}=a_x'(\tau, x')d\tau\label{q4}
\end{align}
Integrating on both sides,
\begin{align}
    &\int\frac{du_x(t, x)}{\left(1-\frac{u_x^2}{c^2}\right)}=\int a_x'(\tau, x')d\tau\nonumber\\
    \text{or\,\,\,\,\,\,\,}&c \tanh^{-1}\left(\frac{u_x}{c}\right)=\int a_x'(\tau, x')d\tau\nonumber\\
   \text{or\,\,\,\,\,\,\,}&u_x=c\tanh\Big[\frac{1}{c}\int a_x'(\tau, x')d\tau\Big] \label{q5}
\end{align}
Further, from equation \eqref{q5} we can obtain as:
\begin{align}
    dx&=c\tanh\Big[\frac{1}{c}\int a_x'(\tau, x')d\tau\Big]\,dt\nonumber\\
   \text{or\,\,\,\,\,\,\,} x&=\int c\tanh\Big[\frac{1}{c}\int a_x'(\tau, x')d\tau\Big]\,dt\label{q6}
\end{align}
 Also we wrote earlier $d\tau=\sqrt{1-\frac{u_x^2}{c^2}}dt$ and the expression for $u_x$ is available in equation \eqref{q5} we can write :
\begin{align}
    t=\int\Big[1-\frac{1}{c^2}\Big\{c\tanh\Big[\frac{1}{c}\int a_x'(\tau, x')d\tau\Big]\Big\}^2\Big]^{-1/2}d\tau\label{w7}
\end{align}
 Introducing  dummy variables $\tau_1$  and $\tau_2$,  in general we can write :
\begin{align}
    t&=\int_0^{\tau}\Big[1-\frac{1}{c^2}\Big\{c\tanh\Big[\frac{1}{c}\int_0^{\tau_1} a_x'(\tau_2, x')d\tau_2\Big]\Big\}^2\Big]^{-1/2}d\tau_1\label{q7}
\end{align}
 Equations \eqref{q6} and \eqref{w7} with appropriate constants of integration will help us to find appropriate  co-ordinate  transformation equations where the acceleration is not constant, but varying with time $(t')$.\\\\
So, in our present case, we take $a^{\prime}_x$ to be a function  of time $(t')$ and  we  shall consider two standard cases: (i) the frame $S^{\prime}$ undergoing Simple Harmonic Motion (SHM) and (ii) the frame $S^{\prime}$ undergoing Uniform Circular Motion (UCM).

\subsection{The  frame  $S^{\prime}$ undergoing SHM in $X-$direction: }
In the first case, we  consider that acceleration $a^{\prime}_x$ is a function of proper time $\tau$ or $(t')$ (we note  $\tau$  and  $t'$  are  identical in our present problem) and represents a simple harmonic motion (SHM). So we may take a displacement $x'=r_0\sin(\omega \tau)$ such that the velocity of the oscillation $u'_x=\frac{dx'}{d\tau}=r_0\omega\cos(\omega\tau)$ and the acceleration $a^{\prime}_x =\frac{d^2x'}{d\tau^2}= -r_0\omega^2 \sin(\omega \tau)$, where $r_0$ is the amplitude of the SHM and $\omega$ is the angular frequency (to   note  $a'_x$  is  co-ordinate  acceleration   and  $\omega$   is  measured  in the $S'$ frame). Therefore, substituting $a_x^{\prime}=-r_0\omega^2 \sin(\omega \tau)$ in equation \eqref{q4} we obtain:

\begin{align}
   &\frac{du_x}{\left(1-\frac{u_x^2}{c^2}\right)}=- r_0\omega^2\sin(\omega \tau)\,d\tau\label{x4}
   \end{align}
Now, integrating {\color{magenta}\bf on} both side of the equation \eqref{x4} we obtain as:
   \begin{align}
   &\int\frac{du_x}{\left(1-\frac{u_x^2}{c^2}\right)}=- r_0\omega^2\int\sin(\omega \tau)\,d\tau\nonumber\\
   \text{or\,\,\,\,\,}&c\,\tanh^{-1}\left(\frac{u_x}{c}\right)= r_0\omega\cos(\omega \tau)+K_1\nonumber\\
   \text{or\,\,\,\,\,\,} &u_x=c\,\tanh\left[\frac{r_0\omega}{c}\cos(\omega \tau)+\frac{ K_1}{c}\right]\label{Bx19}
\end{align}
where $K_1$ is an integration constant. Now, we consider that the displacement vector (of the  $S'$ frame)  is  purely  oscillatory in nature,  so the  three velocity  of  the $S'$  frame  with  respect  to the inertial observer ($S$ frame)  should  be  also oscillatory  in nature  with  zero  base line.  Actually  one  can show  $\tanh{\left[\frac{r_0\omega}{c}\cos(\omega \tau)+\frac{ K_1}{c}\right]}$  function is  oscillatory in nature  with  baseline  which depends upon our choice of the constant $K_1$.  But if  we set $K_1=0$, then  the  baseline  also vanishes  to zero.  So  we  can safely assume  $K_1$ to be equal to  $0$.  \\\\
Therefore, from equation \eqref{Bx19} putting $K_1=0$ we can write as follows: 
\begin{align}
   u_x&=\frac{d x}{dt}=c\,\tanh\left[\frac{r_0\omega}{c}\cos(\omega \tau)\right]\label{qx11}\\
    \text{or\,\,\,\,\,} dx&= c\,\tanh\left[\frac{r_0\omega}{c}\cos(\omega \tau)\right]dt\label{x7}
    \end{align}
So, similarly as before, using the relation : $dt=\Big(1-\frac{u_x^2}{c^2}\Big)^{-1/2}\,d\tau$ and putting the expression for $u_x$ from equation \eqref{qx11} we can re-write equation \eqref{x7} as:
    \begin{align}
    dx&= c\,\tanh\left[\frac{r_0\omega}{c}\Big\{\cos(\omega \tau)\right]\frac{d\tau}{\sqrt{1-\frac{u_x^2}{c^2}}}\nonumber\\
   \text{or\,\,\,\,\,} x&= \int\frac{c\,\tanh\left[\frac{r_0\omega}{c}\Big\{\cos(
   \omega \tau)\right]}{\sqrt{1-\tanh^2\left[\frac{r_0\omega}{c}\cos(\omega \tau)\right]}}d\tau\label{x9}
\end{align}
The right hand side (RHS) of the above equation \eqref{x9} is analytically not integrable. However, as in this equation $r_0\omega$ represents the maximum value of the linear velocity of the frame executing simple harmonic motion (SHM), and for all  practical  purposes  we assume this velocity will be much smaller than the speed of light $(c)$. Thus, assuming $\frac{r_0\,\omega}{c}<<1$, we can write: $\frac{r_0\omega}{c}\cos(\omega \tau)\sim 0$ and hence the right hand side (RHS) of equation \eqref{x9} can be expanded in Taylor series around zero (which  is also called Maclaurin Series) in the power of $r_0\omega/c$. Now,  expanding the integrand  in the  RHS  of  equation \eqref{x9}  and  neglecting the terms higher than {\bf $(r_0\omega/c)^3$} in order, we can obtain from the above equation \eqref{x9} as below:
\begin{align}
     x&= c\,\int\Bigg[0+\frac{r_0\omega}{c}\cos(\omega \tau)+\frac{r_0^2\omega^2}{2c^2}\cdot 0+\frac{r_0^3\omega^3}{6c^3}\cos^3(\omega \tau)\Bigg]d\tau\nonumber\\
     &= r_0\sin(\omega \tau)+\frac{r_0^3\omega^2}{72c^2}\Bigg\{9\sin (\omega\tau )+\sin (3 \omega\tau )\Bigg\}+K_2\label{Bx20}
\end{align}
where $K_2$ is a constant of integration.\\\\
Additionally, since the relation between the proper time ($\tau$) and the coordinate time $(t)$ can be written as: $dt=\Big\{1-\frac{u_x^2}{c^2}\Big\}^{-1/2}\,d\tau$,  hence using equation \eqref{qx11} we can calculate the transformation relation for the time $t$ in case of SHM in $X-$direction as follows:
\begin{align}
    &\int dt=\int\Big\{1-\frac{u_x^2}{c^2}\Big\}^{-1/2}\,d\tau\nonumber\\
   \text{or\,\,\,\,\,\,\,\,\,\,\,\,}&t =\int\left[1-\tanh^2\left[\frac{r_0\omega}{c}\cos(\omega \tau)\right]\right]^{-1/2}d\tau\label{Bqx22}
\end{align}
Again, since as stated above, $r_0\omega/c<<1$ for all practical purposes, we have $\frac{r_0\omega}{c}\cos(\omega \tau)\sim0$, and hence, we can take an approximation by neglecting the terms higher than {\bf$(r_0\omega/c)^3$} as before. So as  before,  expanding the  integrand  in the RHS  of equation \eqref{Bqx22}  in Taylor  series  and utilizing the above  approximation we can obtain from equation \eqref{Bqx22} as:
\begin{align}
    t&\approx \int\left[1+\frac{r_0\omega}{c}\cdot 0+\frac{r_0^2\omega^2}{2c^2}\cos^2(\omega \tau)+\frac{r_0^3\omega^3}{6c^3}\cdot 0\right]d\tau\nonumber\\
    &=\int\left[1+\frac{r_0^2\omega^2}{2c^2}\cos^2(\omega \tau)\right]d\tau\nonumber\\
    &=\tau+\frac{r_0^2\omega}{8c^2}\Bigg\{2 \omega \tau+\sin (2 \omega\tau )\Bigg\}+K_3\label{rxx12}
\end{align}
where $K_3$ is a constant of integration.\\\\
Equations \eqref{Bx20} and \eqref{rxx12} represent the world-line of a particle undergoing time varying acceleration, in terms of its proper time ($\tau$). And using \eqref{Bx20} and \eqref{rxx12}  we can obtain as:
\begin{align}
    x^2-(ct)^2=\mathcal{F}(\tau)\label{Q17}
\end{align}
where $\mathcal{F}(\tau)=\Big[r_0\sin(\omega \tau)+\frac{r_0^3\omega^2}{72c^2}\Big\{9\sin (\omega\tau )+\sin (3 \omega\tau )\Big\}+K_2\Big]^2-\Big[c\tau+\frac{r_0^2\omega}{8c}\Big\{2 \omega \tau+\sin (2 \omega\tau )\Big\}+cK_3\Big]^2$.
Thus  for  a given  $\tau$,  the quantity $a'_x$  can be treated  as  constant,  so using the expression $a_x'=-r_0\omega^2\sin(\omega\tau)$, we can rewrite the function $\mathcal{F}$ as follows:
\begin{align}
    \mathcal{F}&=\Bigg[-\frac{a'_x}{\omega^2}+\frac{r_0^3\omega^2}{72c^2}\left\{-\frac{9a'_x}{r_0\omega^2}-4\left(\frac{a'_x}{r_0\omega^2}\right)^3+3\left(\frac{a'_x}{r_0\omega^2}\right)\right\}+K_2\Bigg]^2\nonumber\\
   &-\Bigg[\frac{c}{\omega}\sin^{-1}\left(-\frac{a'_x}{r_0\omega^2}\right)\left(1+\frac{r_0^2\omega^2}{4c^2}\right)-\frac{r_0}{4c}\frac{a'_x}{\omega}\sqrt{1-\left(\frac{a'_x}{r_0\omega^2}\right)^2}+cK_3\Bigg]^2
\end{align}
which is a  constant  for  a given $\tau$. So, the world-line of a particle with constant acceleration is a hyperbola \cite{wRindler3, Sanchez}. This  fact  verifies  that the  calculations  carried  out  in the  present work  upto  this stage, satisfies  the limiting  conditions of  Rindler's Hyperbola. \\\\
Now, as mentioned above, the frame $S'$ is an inertial frame which instantaneously accompanies the particle undergoing time varying  acceleration and it is evident from the calculations shown in Appendix \ref{AppA} that, the frame $S'$ is a Fermi-Walker transported frame \cite{misner, padmanbhan, kajari_2009}. So, in this situation, the time ($\tau$)  measured by the particle   is equal to the coordinate time ($t'$) of the $S'$ frame.     \\\\
Therefore, in the present context, as $\tau$ can be the time in the frame executing SHM,  
so we may replace $\tau$ by $t'$,  where $t'$ represents the time as measured from the  accelerated frame (as discussed just before equation \eqref{q4}). Therefore, substituting $\tau=t'$ we can obtain from equations \eqref{Bx20} and \eqref{rxx12} as follows:
\begin{align}
x &= r_0\sin(\omega t') +\frac{r_0^3\omega^2}{72c^2}\Bigg[9\sin (\omega t' )+\sin (3 \omega t' )\Bigg]+K_2
    \label{Bx23}\\
   \text{and\,\,\,\,\,\,} t &=t'+\frac{r_0^2\omega}{8c^2}\Big\{2 \omega  t'+\sin (2 \omega t' )\Big\}+K_3\label{BB28}
\end{align}
{\bf Limiting conditions:}
\begin{enumerate}
    \item [(i)]In the initial condition, if the angular frequency of the oscillating frame is equal to zero, i.e. if $\omega=0$ then the coordinates in both frames will coincide, i.e. $x=x'$. So, using that condition, we obtain from equation \eqref{Bx23} as: $K_2=x'$.

\item[(ii)] On the other hand, if there is no oscillation i.e. $\omega=0$ then the time in both frames $S$ and $S'$ should be same, i.e. $t=t'$. So, using that condition we can obtain the constant of integration from equation \eqref{BB28} as: $K_3=0$.
\end{enumerate}
Therefore, putting $K_2=x'$ and $K_3=0$ in equations \eqref{Bx23} and \eqref{BB28} we obtain the coordinate transformation relation for  a frame $S'(t', x', y', z')$  undergoing simple harmonic motion (SHM) in the $X-$direction with respect to an inertial frame $S(t, x, y, z)$ as follows:
    \begin{align}\label{q37}
     t&=t'+\frac{r_0^2\omega}{8c^2}\Big\{2 \omega  t'+\sin (2 \omega t' )\Big\}\\
    x&=x'+r_0\sin(\omega t') +\frac{r_0^3\omega^2}{72c^2}\Big\{9\sin (\omega t' )+\sin (3 \omega t' )\Big\} \label{n18}\\
     y&=y'\\
    z&=z'\label{Eq23}
\end{align}
As mentioned above, the frame $S'$ is essentially a Fermi-Walker transported frame, and locally coincides with an inertial frame along the observer's world-line. So, the frame $S'$ is locally/ instantaneously inertial and preserve the Minkowski metric. So, the transformation relations \eqref{q37}--\eqref{Eq23} are locally Lorentz-type \cite{wRindler3, misner}.\\\\
In the limit of $\omega\longrightarrow 0$ we obtain from equations \eqref{q37} and \eqref{n18}:
\begin{align}
    &\lim_{\omega\to 0}t=t'\label{BBx27}\\
    \text{and\,\,\,\,}&\lim_{\omega\to 0}x=x'\label{BBx28}
\end{align}
Equations \eqref{BBx27} and \eqref{BBx28} above confirm the limiting condition when there is no oscillation $( \omega =0)$. \\\\
Now, in order  to find a relation between time  periods in the two frames we  proceed  as follows. We consider that $T$ and $T'$ are the time periods as observed  from the frames $S$ and $S'$ respectively. From equation \eqref{q37} we can write as:
\begin{align}
    dt&=\Big[1+\frac{r_0^2\omega}{8c^2}\Big\{2\omega+2\omega \cos(2\omega t')\Big\}\Big]dt'\label{qX25a}
\end{align}
Then we integrate both sides of equation \eqref{qX25a},  with limit  $0$ to $T$,  for the unprimed and $0$ to $T'$,  for the primed frame. i.e.
\begin{align}
    \int_0^T dt&=\int_0^{T'}\Big[1+\frac{r_0^2\omega}{8c^2}\Big\{2\omega+2\omega \cos(2\omega t')\Big\}\Big]dt'\nonumber\\
 \text{or\hspace{1.5cm}}   T&=T'+\frac{r_0^2\omega}{8c^2}\Big\{2 \omega  T'+\sin (2 \omega T' )\Big\}\label{w26}
\end{align}
Now, we can clearly see that, $T'$ is the time period as observed  by  an observer sitting in the accelerated frame itself.  So, it  can be identified as  the same time period  as the  non-relativistic  one.  We  denote this  non-relativistic  time period  as $T_0$  and  we can write $T'=T_0$. Therefore, putting $T'=T_0=2\pi/\omega$, we obtain from the above equation \eqref{w26} as follows:
\begin{align}
    \frac{T}{T_0}=1+\frac{1}{4}\frac{\omega^2}{c^2}r_0^2\label{Qx25}
\end{align}
Equation \eqref{Qx25} is almost  identical  to the approximate  expressions  obtained  by other authors \cite{10.1119/1.17513, Fujiwara_2018, zarmi2023}:
\begin{align}
    \frac{T}{T_0}\approx1+\frac{3}{16}\frac{\omega^2}{c^2}r_0^2\label{Qxy27}
\end{align}
Comparing \eqref{Qx25}  and \eqref{Qxy27},  one can  say that,  we  obtained  $(1/4)$  as the  co-effecient of the second term in RHS, corresponding to  $(3/16)$  for the  same  quantity as  obtained  by other authors \cite{10.1119/1.17513, Fujiwara_2018, zarmi2023}.\\\\
This  fact  once  again  confirms  that  all our past calculations based  on which  we  finally obtained the  time periods (in the above),   match   with  the same set of  results obtained by other authors in past.

\subsection{The  frame  $S^{\prime}$ undergoing SHM in $Y-$direction with initial phase of $\pi/2$ radians: }
The circular motion of an object orbiting on the $XY$ plane is best described by two simultaneous SHMs (simple harmonic motions) of same amplitudes in $X$ and $Y$ directions separated by a phase angle of $\pi/2$ radians. 
As we want the $X$ and $Y$ components together to represent a circular motion (with radius $r_0$), we consider a displacement $y'=r_0\cos(\omega\tau)$ of oscillation in $Y-$direction that is separated by a phase angle of $\pi/2$ radians from $x'$ such that the velocity $u'_y=-r_0\omega \sin(\omega \tau)$ and the acceleration $a'_y=-r_0\omega^2 \cos(\omega \tau)$ . \\\\
Therefore, using a similar approach to that in the case of SHM in $X-$direction, we can also obtain the transformation relation of the time and the spatial coordinates for SHM in $Y-$direction. As in the present case, the motion is along the $Y-$direction, therefore coordinate transformation relations can be calculated by simply replacing  $a_x$, $u_x$ and $a_x'$ respectively with $a_y=\frac{du_y}{dt}$, $u_y$ and $a'_y=-r_0\omega^2 \cos(\omega \tau)$ in equation \eqref{q4},
where $u_y$ is the $Y$- component of the three velocity. So, we can write as earlier for SHM in $Y$-direction as follows:
\begin{align}
   &\frac{du_y}{\left(1-\frac{u_y^2}{c^2}\right)}=- r_0\omega^2 \cos(\omega \tau)\,d\tau\label{By30}
\end{align}
Now, integrating both sides of the equation \eqref{By30}, we obtain :
\begin{align}
&\int\frac{du_y}{\left(1-\frac{u_y^2}{c^2}\right)}=-r_0\omega^2 \int \cos(\omega \tau)\,d\tau\nonumber\\
    \text{or\,\,\,\,\,}&c\,\tanh^{-1}\left(\frac{u_y}{c}\right)=-r_0\omega\sin(\omega \tau)+K_4\nonumber\\
    \text{or\,\,\,\,\,} &u_y=c\,\tanh\Bigg[-\frac{r_0\omega}{c}\sin(\omega \tau)+\frac{K_4}{c}\Bigg]\label{By31}
\end{align}
where $K_4$ is an integration constant. Now, again as before, we consider that the displacement vector (of the  $S'$ frame) is  purely  oscillatory in nature,  so the  three velocity  of  the $S'$  frame  with  respect  to the inertial observer ($S$ frame)  should  be  also oscillatory  in nature  with  zero  base line.  Actually,  one  can show that the function $\tanh{\left[-\frac{r_0\omega}{c}\sin(\omega \tau)+\frac{ K_4}{c}\right]}$ is  oscillatory in nature with a baseline which depends upon our choice of the constant $K_4$.  But if  we set $K_4=0$, then  the  baseline  also vanishes  to zero.  So  we  can safely assume  $K_4$ to be equal to  $0$. \\\\ 
Therefore, now equation \eqref{By31} can be written by putting $K_4=0$ as follows: 
\begin{align}
   u_y &=\frac{d y}{dt}=c\,\tanh\Bigg[-\frac{r_0\omega}{c}\sin(\omega \tau)\Bigg]\label{Qxxx29}\\
    \text{or\,\,\,\,\,\,\,}dy&=c\,\tanh\Bigg[-\frac{r_0\omega}{c}\sin(\omega \tau)\Bigg]dt\label{x25}
    \end{align}
Then using the relation $dt=\Big(1-\frac{u_y^2}{c^2}\Big)^{-1/2}\,d\tau$ for SHM in $Y-$direction and putting the expression for $u_y$ from equation \eqref{Qxxx29}, we integrate equation \eqref{x25} as follows:
    \begin{align}
     y&=\int c\,\tanh\Bigg[-\frac{r_0\omega}{c}\sin(\omega \tau)\Bigg]\frac{d\tau}{\sqrt{1-\frac{u_y^2}{c^2}}}\nonumber\\
    &=\int\frac{c\,\tanh\Bigg[-\frac{r_0\omega}{c}\sin(\omega \tau)\Bigg]}{\sqrt{1-\tanh^2\Bigg[-\frac{r_0\omega}{c}\sin(\omega \tau)\Bigg]}}d\tau\label{x27}
\end{align}
The right hand side of equation \eqref{x27} is not integrable. However, as stated before, assuming $r_0\omega/c<<1$, we can write $\frac{r_0\omega}{c}\sin(\omega \tau)\sim 0$ and expand the integrand of the RHS of equation \eqref{x27} in Taylor series in the power of $(r_0\omega/c)$. So, expanding the RHS of equation \eqref{x27} and neglecting the terms higher than $(r_0\omega/c)^3$ we obtain as follows:
\begin{align}
    y&\approx c\int\Big[0+\frac{r_0\omega}{c}\Big\{-\sin(\omega \tau)\Big\}+\frac{r_0^2\omega^2}{2c^2}\cdot 0+\frac{r_0^3\omega^3}{6c^3}\Big\{-\sin(\omega \tau)\Big\}^3\Big]d\tau\nonumber\\
    &=r_0\cos(\omega \tau)+\frac{r_0^3\omega^2}{72c^2}\Bigg[9 \cos (\omega\tau )-\cos (3 \omega\tau )\Bigg]+K_5\label{BBq34}
\end{align}
where $K_5$ is a constant of integration.\\\\
Additionally, as calculated earlier, the transformation relation of time $t$ for SHM in $Y-$direction can be obtained by using the equation \eqref{Qxxx29} as below:
\begin{align}
     &\int dt=\int\Big\{1-\frac{u_y^2}{c^2}\Big\}^{-1/2}\,d\tau\nonumber\\
     \text{or\,\,\,\,\,\,\,\,\,\,}&t=\int\left[1-\tanh^2\left[\frac{r_0\omega}{c}\Big\{-\sin(\omega \tau)\Big\}\right]\right]^{-1/2}d\tau\label{ty}
\end{align}
Now since $r_0\omega/c<<1$, we can take an approximation as before and therefore we can write from equation \eqref{ty} as follows:
\begin{align}
    t&\approx\int\left[1+\frac{r_0\omega}{c}\cdot 0+\frac{r_0^2\omega^2}{2c^2}\Big\{-\sin(\omega \tau)\Big\}^2+\frac{r_0^3\omega^3}{6c^3}\cdot 0\right]d\tau\nonumber\\
   \text{or\,\,\,\,   } t&\approx \tau+\frac{r_0^2 \omega  }{8 c^2}\Big\{2 \omega \tau-\sin (2 \omega\tau )\Big\}+K_6\label{rxx28}
\end{align}
where $K_6$ is a constant of integration.\\\\
Now, as discussed in the above section 2.1, the frame $S'$ is an inertial frame that is instantaneously accompanies the particle undergoing time varying translational acceleration and it can be shown as a Fermi-Walker transported frame \cite{misner, padmanbhan, kajari_2009} (see Appendix \ref{AppA}). So, under the same circumstance, the time $\tau$ measured by the particle (or observer) is equal to the coordinate time ($t'$) of the $S'$ frame.     \\\\
So, in a similar way, since $\tau$ can be the time in the frame executing SHM, we can replace $\tau$ by $t'$  (like was done before for the $X-$component), where $t'$ represents the time measured from the  accelerated frame.  So, substituting $\tau=t'$ we obtain from equations \eqref{BBq34} and \eqref{rxx28} as follows:
\begin{align}
y&=r_0\cos(\omega t')+\frac{r_0^3\omega^2}{72c^2}\Bigg[9 \cos (\omega t')-\cos (3 \omega t' )\Bigg]+K_5 \label{By35}\\
    t&= t'+\frac{r_0^2 \omega  }{8 c^2}\Big\{2 \omega t'-\sin (2 \omega t' )\Big\}+K_6\label{BBy36}
\end{align}
{\bf Limiting conditions:}
\begin{enumerate}
    \item [(i)]In the initial condition, since the SHM in the present case is a cosine function of $t'$ and $\omega$, so if the angular frequency of the oscillating frame is equal to zero, then the coordinates in both frames will be related as: $y=r_0+y'$. So, using that condition, we obtain the integration constant from equation \eqref{By35} as: $K_5=y'$.\\
 
\item[(ii)] On the other hand, if there is no oscillation, i.e. $\omega=0$, then the times in both frames $S$ and $S'$ should be the same, i.e. $t=t'$. So, using that condition, we can obtain the constant of integration from equation \eqref{BBy36} as: $K_6=0$.
\end{enumerate}
Therefore, putting $K_5=y'$ and $K_6=0$ in equations \eqref{By35} and \eqref{BBy36} we obtain the coordinate transformation relation for  a frame $S'(t', x', y', z')$  undergoing simple harmonic motion (SHM) in $Y-$direction with respect to an inertial frame $S(t, x, y, z)$ as follows:
    \begin{align}
     t&=t'+\frac{r_0^2\omega}{8c^2}\Big\{2 \omega  t'-\sin (2 \omega t' )\Big\}\label{Eq37}\\
    x&=x'\\
     y&=y'+r_0\cos(\omega t') +\frac{r_0^3\omega^2}{72c^2}\Big\{9\cos (\omega t' )-\cos (3 \omega t')\Big\}\label{Eq39}\\
    z&=z'\label{Eq40}
\end{align}
In the limiting condition of $\omega\longrightarrow 0$ the equations \eqref{Eq37} and \eqref{Eq39} become:
\begin{align}
    &\lim_{\omega\to 0}t=t'_y\label{bc40}\\
    \text{and\,\,\,\,}&\lim_{\omega\to 0}y=y'+r_0\label{bc41}
\end{align}
The above equations \eqref{bc40} and \eqref{bc41} confirm the limiting condition when there is no oscillation (i.e. $\omega=0$).

\section{Transformation relations for the frame moving in a circular path}
\begin{figure}[hpt]
\centering
\minipage{0.5\textwidth}
\includegraphics[width=\linewidth]{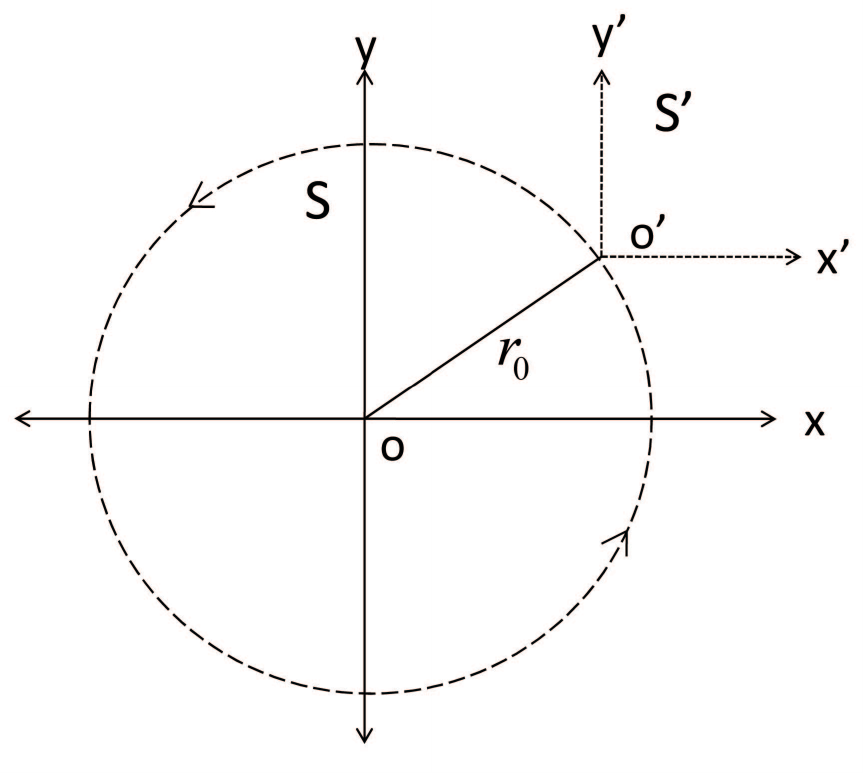}
\caption{\textit{ Schematic of $S'$ frame moving in a circular path on the $XY$ plane.}}
\label{Fig:Qm1}
\endminipage
\end{figure}

As stated above, our objective is to obtain a transformation relation between an inertial frame and a frame moving in a circular path on the $XY$ plane. So we now consider that the frame $S'$ is moving in a circular path with radius $r_0$ where the respective coordinates of two frames are parallel to each other as shown in Fig. \ref{Fig:Qm1}.
 Since the circular motion of an object orbiting on the $XY$ plane can be described by two simultaneous SHMs of the same amplitudes in the $X$ and $Y$ directions separated by a phase angle of $\pi/2$ radians, so the transformation relations of the $X$ and $Y$ coordinates between the inertial frame and the rotating frame can be obtained from equations \eqref{n18} and \eqref{Eq39}; while the $Z$ coordinate will remain unaffected. However, in order to obtain the transformation of the time coordinate in the present case, we used the standard relation between the time coordinate $(t)$ and the proper time $(\tau)$ as follows:
\begin{align}
    t=\int\Big[1-\frac{u^2}{c^2}\Big]^{-1/2}d\tau\label{Eq43}
\end{align}
where $u$ is the velocity on the $XY$ plane in the $S$ frame that can be written in terms of its mutually perpendicular components using equations \eqref{qx11} and \eqref{Qxxx29} as below:
\begin{align}
    u^2&=u^2_x+u^2_y\nonumber\\
    &=c^2\tanh^2\Big\{\frac{r_0\omega}{c}\cos(\omega\tau)\Big\}+c^2\tanh^2\Big\{-\frac{r_0\omega}{c}\sin(\omega\tau)\Big\}\label{Eq44}
\end{align}
Therefore, using equation \eqref{Eq44} into equation \eqref{Eq43} we obtain as follows:
\begin{align}
    t&=\int\Big[1-\Big\{\tanh^2\Big\{\frac{r_0\omega}{c}\cos(\omega\tau)\Big\}+\tanh^2\Big\{-\frac{r_0\omega}{c}\sin(\omega\tau)\Big\}\Big\}\Big]^{-1/2}d\tau\label{EqW43}
\end{align}
Then, as mentioned before, assuming $r_0\omega/c<<1$ we expand the RHS of equation \eqref{EqW43} in Taylor series in the power of $r_0\omega/c$ and neglecting the terms higher than $(r_0\omega/c)^3$ we obtain as follows:
\begin{align}
    t&\approx\int\Big[1+\frac{r_0\omega}{c}\cdot0+\frac{r_0^2\omega^2}{2c^2}\cdot \left( \sin ^2( \omega\tau )+ \cos ^2( \omega\tau )\right)+\frac{r_0^3\omega^3}{6c^3}\cdot 0\Big]d\tau\nonumber\\
    &=\int\Bigg[1+\frac{r_0^2\omega^2}{2c^2}\Bigg]d\tau\nonumber\\
&=\tau+\frac{r_0^2\omega^2}{2c^2}\cdot \tau+K_7\label{EQw44}
\end{align}
where $K_7$ is an integration constant.\\\\
In the present case, the particle is undergoing uniform circular motion and  $\tau$ is the time measured by that  particle. Therefore,  by defining the $S'$ frame moving along with the particle in a circular path where $t'$ represents the coordinate time in that frame, we can write $\tau=t'$ \cite{misner}.\\\\
Therefore, as before, the proper time $\tau$ in the present case can be the time $(t')$ in the frame ($S'$) that is moving in a circular path, and hence the transformation equation for time \eqref{EQw44} can be written as:
\begin{align}
    t=t'\Bigg\{1+\frac{r_0^2\omega^2}{2c^2}\Bigg\}+K_7\label{Eq47}
\end{align}
On the other hand, in the initial condition, if there is no oscillation, i.e., if $\omega=0$ then $t=t'$ and hence the integration constant $K_7$ must be zero. 
Therefore, the coordinate transformation relations for a frame $S'$ orbiting in a circular path can be written from equations \eqref{n18}, \eqref{Eq39} and  \eqref{Eq47} (by putting $K_7=0$) as follows:
\begin{align}
    t&=t'\left\{1+\frac{r_0^2\omega^2}{2c^2}\right\}\label{q45}\\
    x&=x'+r_0\sin(\omega t') +\frac{r_0^3\omega^2}{72c^2}\Big\{9\sin (\omega t' )+\sin (3 \omega t' )\Big\}\label{qx50a}\\
     y&=y'+r_0\cos(\omega t') +\frac{r_0^3\omega^2}{72c^2}\Big\{9\cos (\omega t' )-\cos (3 \omega t' )\Big\}\label{qy51b}\\
    \text{and\,\,\,\,\,}z&=z'\label{q48}
\end{align}
Now, as we have defined that the frame $S'$ is moving along with the observer in a circular path and hence the said transformation relations are locally/instantaneously Lorentz-type transformation.\\\\
Further, we can also write the inverse coordinate transformation relations between an inertial frame and a frame moving in a circular path using equations \eqref{q45}---\eqref{q48} by putting $t'=\lambda t$, where $\lambda=\left\{1+\frac{r_0^2\omega^2}{2c^2}\right\}^{-1}$ as follows:
\begin{align}
    t'&=t\left\{1+\frac{r_0^2\omega^2}{2c^2}\right\}^{-1}=t\,\lambda\\
    x'&=x-r_0\sin(\lambda\omega t) -\frac{r_0^3\omega^2}{72c^2}\Big\{9\sin (\lambda\omega t)+\sin (3 \lambda\omega t )\Big\}\\
     y'&=y-r_0\cos(\lambda\omega t) -\frac{r_0^3\omega^2}{72c^2}\Big\{9\cos (\lambda\omega t)-\cos (3 \lambda\omega t)\Big\}\\
    \text{and\,\,\,\,\,}z'&=z
\end{align}

\section{Comparison with some existing relativistic coordinate transformations}
As mentioned above, in relativity the coordinate transformation relations are important tools that enable us to understand the dynamics of the test particles or an event that occurred in a frame of reference having different states of motion. In literature there exists a good number of works related to the coordinate transformation for non-inertial frames. In 1987, Nelson \cite{nelson1987generalized, 1994JMP....35.6224N} obtained a general coordinate transformation for non-inertial frames having an arbitrary time dependent translational acceleration and angular velocity by generalizing the Lorentz transformation. Ghosh and Sen \cite{doi:10.1139/cjp-2018-1015} discussed about the rotation of  polarization vector  of  an electromagnetic wave and  redshift for  light  emitted from (or  received  by)  a rotating frame, utilizing the  transformation equations as  already  available in  Nelson \cite{nelson1987generalized, 1994JMP....35.6224N}, Mashhoon \cite{MASHHOON1990147} etc.   \\\\
In the present work, we investigated the coordinate transformation relations between an inertial frame and a non-inertial  frame  having non-uniform  acceleration, more specifically the frames executing SHM and UCM. To our knowledge, particularly on the transformation relations for a frame executing SHM, only a few works have been reported in literature. For example, Perepelkin et al. \cite{Perepelkin_2016} discussed about the transformation relation when the velocity of the moving frame is a sinusoidal function of time. Here, the authors obtained a Lorentz-like general coordinate transformation relations for non-inertial frames (see equation (20) and (24) in Perepelkin et al.\cite{Perepelkin_2016}) by using the generalized Galilean transformation. Then the authors considered a time dependent velocity and directly incorporated it into the generalized Lorentz-like transformation. \\\\
On the other hand, in the case of the transformation relation between an inertial frame and a frame moving in a circular path, Nikolic \cite{Nikoli__2000} used Nelson's \cite{nelson1987generalized, 1994JMP....35.6224N} relations to obtain a transformation relation for a rotating frame which is similar to the transformation relations obtained by Mashhoon and Muench \cite{Mashhoon_2002}. \\\\
Additionally, Kipreos and  Balachandran \cite{Kipreos} compared some of the then existing relativistic coordinate transformation relations for rotating frames available in the literature. Here, the authors primarily considered four transformation relations, namely, Post transformation, Franklin transformation, the rotational form of the absolute Lorentz transformation (ALT) and the Langevin metric. And then they compared the theoretical predictions given by those transformation relations with the data including the observations of length contraction, directional time dilation, anisotropic one-way speed of light, isotropic two-way speed of light, and the conventional Sagnac effect. Most  of the  transformation relations  that  the authors  \cite{Kipreos} compared  were  based  on certain assumptions.  However,  in our  present work  we  believe  we extended  Lorentz/ Rindler's  transformation equations in a systematic manner  and  obtained  certain analytical  expressions.  So  we  think  our  work can not  be directly  compared  with the  set of  various similar  work reported  in Kipreos and Balachandran\cite{Kipreos}. The relativistic transformation for a rotating frame obtained by using ALT (following  Franklin \cite{Franklin1922}),  as has been reported in Kipreos and Balachandran\cite{Kipreos} is  probably  the  best  one as  it  matched  most  with the  observational  data. Interestingly, the Langevin relativistic transformation in Cartesian coordinates can be written as follows \cite{Goy_1997}:
\begin{align}
    t=t', \,\,\,\,\,x=x'+r_0 \sin(\omega t'+\phi),\,\,\,\,y=y'+r_0 \cos(\omega t'+\phi)\,\,\,\,\text{and\,\,\,\,}z=z'
\end{align}
which is similar to our results shown in equations \eqref{q45}---\eqref{q48} if we neglect the higher order terms containing $r_0^3\omega^2/(72c^2)$. However, in our coordinate transformation equations, for the frame executing UCM there exist higher order terms containing $r_0^3\omega^2/(72c^2)$ which indicate the relativistic anharmonic effect on the circular motion.

\section{Discussion of results and conclusions}
In this work, we obtained the transformation relations between an inertial frame and a frame executing non-uniform acceleration such as SHM. In a similar way, calculations  were also done to determine the coordinate transformation relations between an inertial frame  and  another frame having uniform circular motion (UCM). These transformation relations can be used to investigate the dynamics of the test particles in the above mentioned non-uniformly accelerated frames. \\\\
Additionally, the results in this paper will be applicable for practical calculations of the relativistic effects, such as in a spacecraft orbiting around the Earth, etc.  Apart from that, these  calculations  will have applications in different branches  of Relativistic  Physics and  more  specifically in Relativistic  Astrophysics and General Relativity. For  example,  materials  in the circumstellar shells  of  pulsating variable  stars undergoing simple  harmonic  motions. In addition we  have  many  binary stars with the  components  moving in circular  orbits around  the  common center of  mass. When    we  study  their  dynamics, or  dynamics  of an electron or  photon (light)  in the gravitational fields  of  such binary stars,  then the  equations  derived  in our present work will  be very useful. \\\\
There  are also interesting  theoretical  works  in   Relativistic  Astrophysics  and  Relativity    in general,  where formulations have  been made under  some  physical situations  for  frames  undergoing  uniform  Rindler  acceleration.  One  such an interesting area is Unruh radiation, where  an observer/detector accelerated through  vacuum  receives  a radiation,  characterized by a   temperature  which is directly proportional to the  proper  acceleration  of the  observer. These results  have  been derived  by various  authors assuming  proper acceleration to be uniform or  constant (for example \cite{10.1143/PTP.88.1, 10.1119/1.1761064}).  Such  results  can now  be  extended  to the  cases  of  SHM and  UCM with  the help of  our  present  work.  Similarly  Xu  et al. \cite{xu2020} had investigated the speedup evolution of the quantum mechanical systems
under the influence of the Unruh effect, where one of
the observers is uniformly accelerated.  So  in our present context,  the  work  by Xu  et al. \cite{xu2020} can be extended  to  non-uniformly  accelerated  cases   like  SHM and  UCM. Fuentes-Schuller  and Mann \cite{Fuentes_Schuller_2005} showed  a state which
is maximally entangled in an inertial frame, becomes less entangled if the observers are relatively
accelerated. This phenomenon, which the authors  derived from the basic Unruh effect, shows that entanglement
is an observer-dependent quantity in non-inertial frames.   Our  present  work  will definitely have  some contributions to offer here,  extending the phenomena to  the  cases  with non-uniform acceleration.  However,  our calculations  as they  are,  can not  be  extended  to the cases where  Unruh radiation takes place within some gravitational field \cite{PhysRevD.14.870}.  This is  because  our transformation equations  are  meant  for flat spacetime and   they essentially follow  Fermi-Walker transport equations.

\section*{Acknowledgments}
The authors thank the faculty members and colleagues of the Department of Physics, Assam University, Silchar (India) for their encouragement and support while doing this work. We are thankful to the anonymous  reviewer for his/her  very  useful suggestions/ comments,  which we believe  have  improved the quality of the paper.
\section*{Declaration}
\textbf{Data Availability:} Since this is a theoretical work, data availability is not applicable and no data has been used or analyzed throughout the work.\\
\textbf{Conflict of interest:} The authors have no relevant financial or non-financial conflict of interests to disclose.\\
\textbf{Funding:}
No funding was received for conducting this study.\\
\textbf{Authors' contribution:} {\it Ranchhaigiri Brahma}--- Methodology, Detail calculations, Formal analysis \& investigation, writing original draft preparation; {\it A.K. Sen}--- Conceptualization, Draft editing, Supervision.

\section*{Appendix}
\appendix
\renewcommand*{\thesection}{\Alph{section}}
\numberwithin{equation}{section}
\numberwithin{table}{section}
\section{Non-uniform acceleration \& Fermi-Walker transported frame}\label{AppA}
A flat spacetime is a spacetime with zero curvature, i.e. where gravity is absent.
In such a spacetime, a reference frame is said to be Fermi-Walker transported along the observer's world-line if it satisfies the following mathematical equation \cite{misner, padmanbhan}:
\begin{align}
    \frac{dV^{\mu}}{d\tau}=\frac{1}{c^2}\left(U^{\mu}A^{\nu}-U^{\nu}A^{\mu}\right)V_{\nu}\label{A1}
\end{align}
where $U^{\mu}$, $A^{\mu}$, $\tau$, and $c$ are four-velocity, four-acceleration, proper time, and speed of light in vacuum, respectively \footnote{The Greek indices ($\mu,\nu, ...$) are used to indicates to run ($0, 1, 2, 3$) that represents $ct$, $x$, $y$ and $z$ co-ordinates respectively.}. $V^{\mu}$ is known as the Fermi-Walker vector, which is a vector along an observer's world-line without rotation, relative to the observer's instantaneous rest frame \cite{padmanbhan}.(A  mathematical  expression for $V^{\mu}$  is given below after equation \eqref{A23} ). The factor $1/c^2$ in equation \eqref{A1} ensures the dimensional consistency of the Fermi-Walker law of transport.\\\\
Now, in order to analyze the Fermi-Walker law of transport in the case of a non-uniformly accelerating frame $S'$, which is instantaneously accompanies the particle undergoing SHM along the $x-$axis, we consider Minkowski coordinates $(ct, x, y, z)$ with the metric sign convention: $\eta_{\mu\nu}=\text{diag}(1,-1,-1,-1)$. So, the world-line of the frame $S'$ in the Minkowski spacetime can be expressed as follows (from equations \eqref{Bx20} and \eqref{rxx12}):
\begin{align}
x&= r_0\sin(\omega \tau)+\frac{r_0^3\omega^2}{72c^2}\Bigg\{9\sin (\omega\tau )+\sin (3 \omega\tau )\Bigg\}+K_2\label{A2}\\
    ct&=c\tau+\frac{r_0^2\omega}{8c}\Bigg\{2 \omega \tau+\sin (2 \omega\tau )\Bigg\}+K_3\label{A3}
\end{align}
where $K_2$ and $K_3$ are constants. Now, we can obtain the four-velocity for the frame $S'$ as follows (based on the procedures as in \cite{padmanbhan, misner}):
\begin{align}
    U^0&=\frac{cdt}{d\tau}=c+\frac{r_0^2\omega^2}{2c}\cos^2(\omega \tau)\label{A4}\\
    U^1&=\frac{dx}{d\tau}= r_0\omega\cos(\omega \tau)+\frac{r_0^3\omega^3}{6c^2}\cos^3(\omega \tau)\label{Ap5}\\
    U^2&=U^3=0\label{A6}
\end{align}
That is,
\begin{align}
    U^{\mu}&=\Big\{c+\frac{r_0^2\omega^2}{2c}\cos^2(\omega \tau),\, r_0\omega\cos(\omega \tau)+\frac{r_0^3\omega^3}{6c^2}\cos^3(\omega \tau),\,0,\,0\Big\}\label{A7}\\
 \text{and\,\,\,\,}    U_{\nu}&=\eta_{\mu\nu}U^{\mu}\nonumber\\
    &=\Big\{c+\frac{r_0^2\omega^2}{2c}\cos^2(\omega \tau),\,- r_0\omega\cos(\omega \tau)-\frac{r_0^3\omega^3}{6c^2}\cos^3(\omega \tau),\,0,\,0\Big\}\label{Ap8}
\end{align}
The norm of the above four-velocity of the frame $S'$ can be written as :
\begin{align}
    U_{\mu}U^{\mu}&=\left\{c+\frac{r_0^2\omega^2}{2c}\cos^2(\omega \tau)\right\}^2- \left\{r_0\omega\cos(\omega \tau)+\frac{r_0^3\omega^3}{6c^2}\cos^3(\omega \tau)\right\}^2\nonumber\\
    &=c^2+r_0^2\omega^2\cos^2(\omega \tau)+\frac{r_0^4\omega^4}{4c^2}\cos^4(\omega \tau)- r_0^2\omega^2\cos^2(\omega \tau)\nonumber\\
    &-\frac{r_0^4\omega^4}{3c^2}\cos^4(\omega \tau)-\frac{r_0^6\omega^6}{36c^4}\cos^6(\omega \tau)\nonumber\\
    &=c^2\left\{1-\frac{r_0^4\omega^4}{48c^4}\cos^4(\omega \tau)-\frac{r_0^6\omega^6}{36c^6}\cos^6(\omega \tau)\right\}\label{A9}
\end{align}
It should be emphasized here that, during our calculations, we assumed $r_0\omega/c<<1$ and neglected the terms higher than $(r_0\omega/c)^3$. So, neglecting the terms higher than $(r_0\omega/c)^3$ we obtain from the above equation \eqref{A9} as follows:

\begin{align}
   U_{\mu}U^{\mu} &=c^2\hspace{1.5cm}\text{and}\hspace{1.5cm}\vert U\vert =c\label{A10}
\end{align}
Now, the four-acceleration of the frame $S'$ can be obtained from equations \eqref{A4}---\eqref{A6} as follows (based on the procedures as in \cite{padmanbhan, misner}):
\begin{align}
    A^0&=\frac{dU^t}{d\tau}=-\frac{r_0^2 \omega ^3}{2c} \sin (2\omega \tau)\\
    A^1&=\frac{dU^x}{d\tau}=-r_0 \omega ^2 \sin (\omega\tau )-\frac{r_0^3 \omega ^4 }{2 c^2}\sin (\omega \tau) \cos ^2(\omega \tau)\\
    A^2&=A^3=0
\end{align}
That is,
\begin{small}
    \begin{align}
    A^{\nu}&=\Big\{-\frac{r_0^2 \omega ^3}{2c} \sin (2\omega \tau),\,\,-r_0 \omega ^2 \sin (\omega\tau )-\frac{r_0^3 \omega ^4 }{2 c^2}\sin (\omega \tau) \cos ^2(\omega \tau),\,0,\,0\Big\}\label{A14}\\
  & \text{and\,\,\,\,}\nonumber\\
   A_{\nu}&=\eta_{\mu\nu}A^{\mu}\nonumber\\
    &=\Big\{-\frac{r_0^2 \omega ^3}{2c} \sin (2\omega \tau),\,\,r_0 \omega ^2 \sin (\omega\tau )+\frac{r_0^3 \omega ^4 }{2 c^2}\sin (\omega \tau) \cos ^2(\omega \tau),\,0,\,0\Big\}\label{A15}
\end{align}
\end{small}\\
The orthogonality of the four-acceleration with respect to four-velocity can be checked as follows:
\begin{align}
    U^{\mu}A_{\mu}&=U^0A_0+U^1A_1\nonumber\\
    &=\left\{c+\frac{r_0^2\omega^2}{2c}\cos^2(\omega \tau)\right\}\left\{-\frac{r_0^2 \omega ^3}{2c} \sin (2\omega \tau)\right\}+\Bigg\{r_0\omega\cos(\omega \tau)\nonumber\\
    &+\frac{r_0^3\omega^3}{6c^2}\cos^3(\omega \tau)\Bigg\}\left\{r_0 \omega ^2 \sin (\omega\tau )+\frac{r_0^3 \omega ^4 }{2 c^2}\sin (\omega \tau) \cos ^2(\omega \tau)\right\}\nonumber\\
    &=-\frac{r_0^2 \omega ^3}{2} \sin (2\omega \tau)-\frac{r_0^4 \omega ^5}{4c^2} \sin (2\omega \tau)\cos^2(\omega\tau)\nonumber\\
    &+r_0^2\omega^3\sin(\omega\tau)\cos(\omega \tau)+\frac{r_0^4\omega^5}{6c^2}\sin(\omega\tau)\cos^3(\omega \tau)\nonumber\\
   & +\frac{r_0^4 \omega ^5 }{2 c^2}\sin (\omega \tau) \cos ^3(\omega \tau)+\frac{r_0^6 \omega ^7 }{12 c^4}\sin (\omega \tau) \cos ^5(\omega \tau)\nonumber\\
   &=\Bigg[\frac{r_0^4\omega^5}{2c^4}\cos^3(\omega\tau)\Big\{-\cos (\omega \tau)+\frac{4}{3}\Big\}+\frac{r_0^6 \omega ^7 }{12 c^6} \cos ^5(\omega \tau)\Bigg]c^2\sin(\omega\tau)\label{A16}
\end{align}
Again as stated before neglecting the terms containing higher than $(r_0\omega/c)^3$ we obtain as:
\begin{align}
    U^{\mu}A_{\mu}=U_{\mu}A^{\mu}=0\label{A17}
\end{align}
which indicates that four-velocity $(U^{\mu})$ and four-acceleration $(A^{\mu})$ in the present context are orthogonal to each other. Additionally, the norm of the four-acceleration can be obtained as follows:
\begin{small}
\begin{align}
    A_{\mu}A^{\mu}&=A_{0}A^{0}+A_{1}A^{1}\nonumber\\
    &=\left\{-\frac{r_0^2 \omega ^3}{c} \sin (\omega \tau) \cos (\omega\tau )\right\}^2- \left\{r_0 \omega ^2 \sin (\omega\tau )+\frac{r_0^3 \omega ^4 }{2 c^2}\sin (\omega \tau) \cos ^2(\omega \tau)\right\}^2\nonumber\\
    &=\frac{r_0^4 \omega ^6}{c^2} \sin^2 (\omega \tau) \cos^2 (\omega\tau )-r_0^2\omega^4\sin^2(\omega\tau)-\frac{r_0^4 \omega ^6 }{ c^2}\sin^2 (\omega \tau) \cos ^2(\omega \tau)\nonumber\\
    &-\frac{r_0^6 \omega ^8 }{4 c^4}\sin^2 (\omega \tau) \cos ^4(\omega \tau)\nonumber\\
    &=\Bigg[-\frac{r_0^2\omega^4}{c^2}-\frac{r_0^6 \omega ^8 }{4 c^6} \cos ^4(\omega \tau)\Bigg]c^2\sin^2(\omega\tau)
\end{align}
\end{small}\\
Again, neglecting the terms containing higher than $(r_0\omega/c)^3$ we obtain:
\begin{align}
    A_{\mu}A^{\mu}=-r_0^2\omega^4\sin^2(\omega\tau)\hspace{1.1cm}\text{and}\hspace{1.1cm}\vert A\vert=r_0\omega^2\sin(\omega\tau)
\end{align}
So, the components of the natural orthogonal tetrad carried by the frame $S'$ can be defined as follows\cite{kajari_2009, padmanbhan}:
\begin{align}
    e^{\mu}_{(0)}&=\frac{U^{\mu}}{\vert U\vert }\\
     e^{\mu}_{(1)}&=\frac{A^{\mu}}{\vert A\vert}=\Big\{-\frac{r \omega }{c} \cos (\omega\tau ),\,\,-1-\frac{r^2 \omega ^2 }{2 c^2}\cos ^2(\omega\tau ),\,0,\,0 \Big\}\label{A21}\\
      e^{\mu}_{(2)}&=\left(0,0,1,0\right)\\
       e^{\mu}_{(3)}&=\left(0,0,0,1\right)\label{A23}
\end{align}
The covariant form of the above tetrads can be obtained by using the transformation relation $e_{(i)\mu}=\eta_{\mu\nu}e^{\mu}_{(i)}$, where $i=0,1,2,3$. Further, since the Fermi-Walker vector is $V^{\mu}=e^{\mu}_{(1)}$ \cite{padmanbhan} and hence the left hand side of the Fermi-Walker transport condition \eqref{A1} can be expressed by using \eqref{A21} as follows:
\begin{align}
   \frac{dV^{\mu}}{d\tau}= \frac{de^{\mu}_{(1)}}{d\tau}&=\Big\{\frac{r \omega ^2 }{c}\sin ( \omega\tau ),\,\frac{r^2 \omega ^3 }{c^2}\sin ( \omega\tau ) \cos (\omega\tau ),\,0,\,0\Big\}\label{Ap24}
\end{align}
Now, using equations \eqref{Ap8} and \eqref{A21} we obtain:
\begin{align}
    U^{\nu}e_{(1)\nu}&=U^{\nu}\frac{A_{\nu}}{\vert A\vert}=0
\end{align}
and using equations \eqref{A15} and \eqref{A21} we obtain as:
\begin{align}
    A^{\nu}e_{(1)\nu}&=\left\{-\frac{r_0^2 \omega ^3}{2c} \sin (2\omega \tau)\right\}\left\{-\frac{r \omega }{c} \cos (\omega\tau )\right\}-\Big\{r_0 \omega ^2 \sin (\omega\tau )\nonumber\\
    &+\frac{r_0^3 \omega ^4 }{2 c^2}\sin (\omega \tau) \cos ^2(\omega \tau)\Big\}\left\{1+\frac{r^2 \omega ^2 }{2 c^2}\cos ^2(\omega\tau )\right\}\nonumber\\
    &=\frac{r_0^3 \omega ^4}{2c^2} \sin (2\omega \tau)\cos(\omega\tau)-r_0 \omega ^2 \sin (\omega\tau )-\frac{r_0^3 \omega ^4 }{2 c^2}\sin (\omega \tau) \cos ^2(\omega \tau)\nonumber\\
    &-\frac{r^3 \omega ^4}{2 c^2} \cos ^2(\omega\tau )\sin(\omega\tau)-\frac{r^5 \omega ^6}{4 c^4}\sin(\omega\tau)\cos^4(\omega\tau)\nonumber\\
    &=-r_0 \omega ^2 \sin (\omega\tau )-\frac{r^5 \omega ^6}{4 c^4}\sin(\omega\tau)\cos^4(\omega\tau)\nonumber\\
    &\approx -r_0 \omega ^2 \sin (\omega\tau )
\end{align}
Therefore, the right hand side of equation \eqref{A1} can be expressed as follows:
\begin{align}
   \frac{1}{c^2}\Big(U^{\nu}A^{\mu}&-U^{\mu}A^{\nu}\Big)e_{(1)\nu}=\frac{1}{c^2}\Big\{A^{\mu}\left(U^{\nu}e_{(1)\nu}\right)-U^{\mu}\left(A^{\nu}e_{(1)\nu}\right)\Big\}\nonumber\\
    &=\frac{1}{c^2}A^{\mu}\left(0\right)-\frac{1}{c^2}U^{\mu}\left\{-r_0\omega^2\sin(\omega\tau)\right\}\nonumber\\
    &=\Big\{\frac{1}{c}+\frac{r_0^2\omega^2}{2c^3}\cos^2(\omega \tau),\, \frac{r_0\omega}{c^2}\cos(\omega \tau)+\frac{r_0^3\omega^3}{6c^4}\cos^3(\omega \tau),\,0,\,0\Big\}\nonumber\\
    &\hspace{4cm}\times\left(r_0\omega^2\sin(\omega\tau)\right)\nonumber\\
    &=\Big\{\frac{r_0\omega^2}{c}\sin(\omega\tau)+\frac{r_0^3\omega^4}{2c^3}\cos^2(\omega \tau)\sin(\omega\tau),\nonumber\\
    &+\frac{r^2_0\omega^3}{c^2}\cos(\omega \tau)\sin(\omega\tau)+\frac{r_0^4\omega^5}{6c^4}\cos^3(\omega \tau)\sin(\omega\tau),\,0,\,0\Big\}\nonumber\\
    &\approx \Big\{\frac{r_0\omega^2}{c}\sin(\omega\tau),\,\,\frac{r^2_0\omega^3}{c^2}\cos(\omega \tau)\sin(\omega\tau),\,0,\,0\Big\} \label{Ap27}
\end{align}
Therefore, from equations \eqref{Ap24} and \eqref{Ap27} we obtain that a frame $S'$ which is instantaneously accompanies the particle undergoing non-uniform acceleration (having world-line \eqref{A2} and \eqref{A3}) is a Fermi-Walker transported frame.\\\\
Similarly, we can verify that the frame $S'$ undergoing SHM along the $y$-axis with initial phase difference of $\pi/2$ radians is also a Fermi-Walker transported frame as follows: \\\\
In the present case, using the world-line represented by equations \eqref{BBq34} and \eqref{rxx28}, we can obtain the four-velocity $(U^{\mu})$ and four-acceleration $(A^{\mu})$ as below:
\begin{small}
\begin{align}
    U^{\mu}&=\Big\{c+\frac{r_0^2\omega^2}{2c}\sin^2(\omega \tau),\,0,\,- r_0\omega\sin(\omega \tau)-\frac{r_0^3\omega^3}{6c^2}\sin^3(\omega \tau),\,0\Big\}\label{Ap28}\\
 \text{and\,\,}\nonumber\\
    A^{\nu}&=\Big\{\frac{r_0^2 \omega ^3}{2c} \sin (2\omega \tau),\,0,\,\,-r_0 \omega ^2 \cos (\omega\tau )-\frac{r_0^3 \omega ^4 }{2 c^2}\sin^2 (\omega \tau) \cos (\omega \tau),\,0\Big\}
\end{align}
\end{small}\\
Furthermore, we calculate the norm of the four-acceleration as below:
\begin{small}
\begin{align}
    A_{\mu}A^{\mu}&=A_{0}A^{0}+A_{2}A^{2}\nonumber\\
    &=\left\{\frac{r_0^2 \omega ^3}{c} \sin (\omega \tau) \cos (\omega\tau )\right\}^2- \left\{r_0 \omega ^2 \cos (\omega\tau )+\frac{r_0^3 \omega ^4 }{2 c^2}\sin^2 (\omega \tau) \cos(\omega \tau)\right\}^2\nonumber\\
    &=\frac{r_0^4 \omega ^6}{c^2} \sin^2 (\omega \tau) \cos^2 (\omega\tau )-r_0^2\omega^4\cos^2(\omega\tau)-\frac{r_0^4 \omega ^6 }{ c^2}\sin^2 (\omega \tau) \cos ^2(\omega \tau)\nonumber\\
    &-\frac{r_0^6 \omega ^8 }{4 c^4}\sin^4 (\omega \tau) \cos ^2(\omega \tau)\nonumber\\
    &=\Bigg[-\frac{r_0^2\omega^4}{c^2}-\frac{r_0^6 \omega ^8 }{4 c^6} \sin ^4(\omega \tau)\Bigg]c^2\cos^2(\omega\tau)
\end{align}
\end{small}\\
Now, neglecting the terms containing higher than $(r_0\omega/c)^3$ as before, we obtain:
\begin{align}
    A_{\mu}A^{\mu}=-r_0^2\omega^4\cos^2(\omega\tau)\hspace{1.1cm}\text{and}\hspace{1.1cm}\vert A\vert=r_0\omega^2\cos(\omega\tau)\label{Ap31}
\end{align}
Additionally, we can check the orthogonality between the four-velocity and four-acceleration as follows:
\begin{align}
    U^{\mu}A_{\mu}&=U^0A_0+U^2A_2\nonumber\\
    &=\left\{c+\frac{r_0^2\omega^2}{2c}\sin^2(\omega \tau)\right\}\left\{\frac{r_0^2 \omega ^3}{2c} \sin (2\omega \tau)\right\}+\Bigg\{-r_0\omega\sin(\omega \tau)\nonumber\\
    &-\frac{r_0^3\omega^3}{6c^2}\sin^3(\omega \tau)\Bigg\}\left\{r_0 \omega ^2 \cos (\omega\tau )+\frac{r_0^3 \omega ^4 }{2 c^2}\sin^2 (\omega \tau) \cos (\omega \tau)\right\}\nonumber\\
    &=\frac{r_0^2 \omega ^3}{2} \sin (2\omega \tau)+\frac{r_0^4 \omega ^5}{4c^2} \sin (2\omega \tau)\sin^2(\omega\tau)\nonumber\\
    &-r_0^2\omega^3\sin(\omega\tau)\cos(\omega \tau)+\frac{r_0^4\omega^5}{6c^2}\sin^3(\omega\tau)\cos(\omega \tau)\nonumber\\
   & -\frac{r_0^4 \omega ^5 }{2 c^2}\sin^3 (\omega \tau) \cos (\omega \tau)-\frac{r_0^6 \omega ^7 }{12 c^4}\sin^5 (\omega \tau) \cos (\omega \tau)\nonumber\\
   &\approx 0\label{Ap32}
\end{align}
So, by defining $V^{\mu}=e_{(1)}^{\mu}=\frac{A^{\mu}}{\vert A\vert}$ we can determine the LHS of the Fermi-Walker transported law \eqref{A1} as:
\begin{align}
   \frac{dV^{\mu}}{d\tau}= \frac{de^{\mu}_{(1)}}{d\tau}&=\Big\{\frac{r \omega ^2 }{c}\cos ( \omega\tau ),\,0,\,-\frac{r^2 \omega ^3 }{c^2}\sin ( \omega\tau ) \cos (\omega\tau ),\,0\Big\}\label{Ap33}
\end{align}
And using equations \eqref{Ap28}, \eqref{Ap31} and \eqref{Ap32} we can calculate the RHS of equation \eqref{A1} as follows:

\begin{align}
   \frac{1}{c^2}\Big(U^{\nu}A^{\mu}&-U^{\mu}A^{\nu}\Big)e_{(1)\nu}=\frac{1}{c^2}\Big\{A^{\mu}\left(U^{\nu}e_{(1)\nu}\right)-U^{\mu}\left(A^{\nu}e_{(1)\nu}\right)\Big\}\nonumber\\
    &=\frac{1}{c^2}A^{\mu}\left(U^{\nu}\frac{A_{\nu}}{\vert A\vert}\right)-\frac{1}{c^2}U^{\mu}\left\{-r_0\omega^2\cos(\omega\tau)\right\}\nonumber\\
    &=\frac{1}{c^2}A^{\mu}\left(0\right)-\frac{1}{c^2}U^{\mu}\left\{-r_0\omega^2\cos(\omega\tau)\right\}\nonumber\\
    &=\Big\{\frac{1}{c}+\frac{r_0^2\omega^2}{2c^3}\sin^2(\omega \tau),\,0,\, -\frac{r_0\omega}{c^2}\sin(\omega \tau)-\frac{r_0^3\omega^3}{6c^4}\sin^3(\omega \tau),\,0\Big\}\nonumber\\
    &\hspace{4cm}\times\left(r_0\omega^2\cos(\omega\tau)\right)\nonumber\\
    &=\Big\{\frac{r_0\omega^2}{c}\cos(\omega\tau)+\frac{r_0^3\omega^4}{2c^3}\cos(\omega \tau)\sin^2(\omega\tau),\,0\nonumber\\
    &-\frac{r^2_0\omega^3}{c^2}\cos(\omega \tau)\sin(\omega\tau)-\frac{r_0^4\omega^5}{6c^4}\cos(\omega \tau)\sin^3(\omega\tau),\,0\Big\}\nonumber\\
    &\approx \Big\{\frac{r_0\omega^2}{c}\cos(\omega\tau),\,0,\,\,-\frac{r^2_0\omega^3}{c^2}\cos(\omega \tau)\sin(\omega\tau),\,0\Big\} \label{Ap34}
\end{align}
From equations \eqref{Ap33} and \eqref{Ap34} we can state that the frame $S'$ undergoing SHM in $y-$direction with initial phase difference of $\pi/2$ radians in the present case is also a Fermi-Waller transported frame.

%\bibliographystyle{unsrt}
%\bibliography{Brahma_Sen_ms}

\end{document}